\documentclass{FrontiersinHarvard}

\usepackage{url,hyperref,lineno,microtype}
\usepackage[onehalfspacing]{setspace}

\def\keyFont{\fontsize{8}{11}\helveticabold }

\def\Authors{Krista Lynne Smith\,$^{1,*}$}
\def\Address{$^{1}$ Mitchell Institute for Fundamental Physics and Astronomy, Texas A\&M University, College Station, TX, USA 
}
\def\corrAuthor{Corresponding Author}

\def\corrEmail{kristalynnesmith@tamu.edu}

\begin{document}
\onecolumn
\firstpage{1}

\title[High-Energy Physics with Multi-Color Space-Based Timing]{Rapid Multi-band Space-Based Optical Timing: Revolutionizing Accretion Physics} 

\author[\firstAuthorLast ]{\Authors} 
\address{} 
\correspondance{} 

\extraAuth{}
\maketitle
\begin{abstract}

Optical timing with rapid, seconds-to-minutes cadences with high photometric precision and gap-free long baselines is necessary for an unambiguous physical picture of accretion phenomena, and is only possible from space. Exoplanet-hunting missions like Kepler and TESS have offered an outstanding new window into detailed jet and accretion physics, but have been severely hampered by incomplete calibration and systematics treatments and, most especially, a monochromatic single wide bandpass. Advances made using Kepler and TESS survey data, when considered alongside detailed, expensive multi-color experiments done from the ground, reveal the enormous potential of a space-based multi-color optical timing mission with a high energy focus. 

\tiny
 \keyFont{ \section{Keywords:} AGN - Active Galactic Nucleus, Blazars, time domain, accretion, black holes} 
\end{abstract}

\section{Introduction}

During its prime mission from 2009 - 2013, the \emph{Kepler} spacecraft fulfilled its brief spectacularly, discovering thousands of extrasolar planets. Its success was due to three major properties of the mission tailored to the search for exoplanet transits: rapid, 30-minute cadence of the observations, a long and uninterrupted $\sim4$ year baseline, and unprecedented photometric precision. The lack of seasonal and diurnal gaps, so often the bane of ground-based monitoring campaigns, enabled a full temporal sampling space. The properties that made the mission such an exemplary planet-finder also raised the potential for it to study variable high-energy phenomena in unexplored parameter spaces as well. A number of investigations into active galactic nuclei (AGN) and other accreting systems took place, but were hampered significantly by instrumental systematics that required complex bespoke reduction techniques often difficult to reproduce. To a large extent, this was due to the non-exoplanet applications requiring more complete calibration than needed for detecting the strictly periodic point-source variation that characterizes exoplanet transits; see \citet{Smith2019} for a complete discussion, and \citet{Howell2020} for a summary of \emph{Kepler} results, including those beyond exoplanets.

The successor to \emph{Kepler}, the Transiting Exoplanet Survey Satellite (TESS) was launched in 2018 with the same high precision, rapid cadence, and monolithic bandpass (with a redder central wavelength) as its predecessor, but with a few key differences, including the release of early-stage full frame images (FFIs) for the entire survey region and a much larger, nearly all-sky survey footprint, with the tradeoff of shorter baselines (27 days to 1 year, depending on ecliptic latitude). From the start, TESS made itself more amenable to non-exoplanet science applications by releasing flexible support software for reduction \citep[e.g. ][ originally designed for K2 data]{Lightkurve2018} and calibration products for the FFIs. Nonetheless, TESS systematics make time domain analysis of stochastically varying sources like AGN quite challenging, especially when stitching light curves across many 27-day sectors. Fortunately, TESS supported user-created reduction software as part of its guest observer program, resulting in a proliferation of pipelines for both generic and highly customized science applications. Still, many works publish light curves that are badly affected by systematics. 

In addition to challenging systematics due to electronics and background light, both missions also have a single white-light bandpass. While an efficient choice for transient and exoplanet searches, this prohibits the vast amount of astrophysics present in cross-band correlations at rapid timescales. Because these investigations require significant resources with existing instruments, almost all experiments at high cadence are done in expensive single object case studies.

\section{Potential High-Energy Applications of Multi-colored Space-based Timing}
\label{sec:ground}

\subsection{Searching for Binary Supermassive Black Holes}
\label{sec:bbh}

A confident census of the separations, mass ratios, and spins of binary supermassive black holes is an important prior for the multimessenger detections of binary inspirals through gravitational waves, both the stochastic background of orbiting binary pairs as seen by pulsar timing arrays like NANOGrav \citep{Agazie2023} and in future signals expected from the upcoming LISA observatory \citep{Amaro2017}. One potential signal is periodicity in AGN light curves due to Doppler boosting of  an orbiting pair, periodic accretion episodes in the mini-disks, or precessing jets \citep[e.g., ][]{Dorazio2015,Ryan2017,Charisi2018,Combi2022}; observed candidates of each of these have been put forth in varying numbers \citep[e.g., ][]{Graham2015,Liu2016,Britzen2018,Liao2021}. However, the red noise nature of accretion disk and jet variability can lead to false positives in searches for binaries, which are seriously exacerbated by gaps in light curves. As the following examples demonstrate, far less ambiguous signatures of binary AGN are possible with high-cadence, uninterrupted monitoring, and adding color information increases the confidence even more. 

Self-lensing flares, wherein a foreground massive black hole gravitationally lenses the emission from the mini-disk around the background hole in an orbiting pair, is a periodic signal that occurs at a predictable phase of a doppler-boosted periodic oscillation, and is not degenerate with other physical explanations \citep{Dorazio2018}. Only one such candidate exists currently, and it is from Kepler data, which was capable of capturing the approximately week-long flare event with very high significance \citep{Hu2020}. 

As shown by \citet{Dorazio2018}, when the background accretion disk is near enough to be lensed as a source of finite size, multi-band light curves would provide a very sensitive probe of the structure of that accretion flow, a totally unique observable of the fueling of binaries. If the background disk is a point source, lensing is, as usual, achromatic; this would make the self-lensing flare completely distinct from other AGN flares. 

The technique of ``varstrometry" has recently been used with \emph{Gaia} data to locate binary AGN candidates (at much larger separations than the self-lensing or periodicity techniques) using astrometric noise: two AGN will vary independently, so even when they are convolved in a low-resolution image, the image's centroid will shift as one or the other AGN becomes brighter \citep{Liu2015, Shen2019, Hwang2020}. This method has successfully recovered a large number of known close binaries, and led to the discovery of many new binaries \citep[e.g., ][]{Chen2022}, although the technique is also useful for discovering lensed images of single quasars \citep[e.g., ][]{Shen2021,Springer2021}. Rapid monitoring will allow for tighter constraints on astrometric noise, as well as noise at different variability timescales, and reveal a wider population of astrometrically-varying binary candidates. Furthermore, adding colors to the ``varstrometry" method widens the discovery space even further; as the centroid shifts, so will the peak wavelength at which the convolved source is varying. This will lead to distinct centroids in different wavebands, as shown by \citet{Liu2015} and \citet{LiuY2016}. The achromaticity of gravitational lensing also suggests that multi-color varstrometry is likely to be able to differentiate between binary AGN pairs and lensing as the underlying cause of the shifting light center.

\subsection{Relativistic Jet Physics in Blazars}
\label{sec:blazars}

Blazars vary significantly at a huge variety of timescales, from a few minutes to a few years, in every waveband in which they have been studied. A vast literature exists focused on simultaneous monitoring of blazars across the full electormagnetic spectrum, with the most frequent goal being the determination of the origin of the two peaks of the blazar spectral energy distribution (SED). The lower-energy peak can span the radio to UV (so-called ``low-synchrotron peak" or LSP blazars) or X-ray (``high-synchrotron peak"; HSP) wavebands, while the higher energy peak extends blue-ward from a trough after the low energy peak, including the gamma ray spectral region. The physical origin of the low-energy peak is well understood as synchrotron emission from electrons in the relativistic jet powered by the supermassive black hole. The gamma ray emission in the high-energy peak could come from leptonic processes like Compton-upscattered thermal photons from the accretion disk or obscuring torus \citep[external Compton models; ][]{Sikora1993} or upscattered synchrotron photons from the jet itself \citep[self-Compton models; ][]{Mastichiadis1997}. Alternatively, the gamma ray emission could come from hadronic processes such as pion decay or proton- synchrotron radiation \citep[e.g., ][]{Bottcher2009, Reimer2012}. Self-compton models imply that a strong correlation between variations in the lower and higher energy peaks would be expected, while hadronic models might suggest a weak or no correlation. 

Within the optical band, multi-color monitoring has been used in a large number of ground-based surveys with semi-daily to weekly sampling over many years to study the physics of blazar flares. \citet{Bonning2012} showed convincing evidence for a shift in the low energy peak of the blazar 3C~454.3 over a 24 hour flaring period in 2009 using multi-band optical-IR SMARTS monitoring, and reported a redder-when-brighter trend in flat-spectrum radio quasars (FSRQs). Recent ground-based studies have found a bluer-when-brighter trend in many blazars, as well as using the rapid minutes-scale variability to determine upper limits on black hole masses, a very challenging quantity to measure in beamed targets like these \citep{Chang2023,Li2024}. Other studies have found that whether a source becomes bluer or redder when brightening depends on its optical spectral classification as an FSRQ or a BL Lac object \citep{Zhang2023}. The amount and color lag properties of so-called ``microvariability" on minutes-to-hours timescales can be used to study the uniformity of particle flow within the jets and meaningfully constrain the amount of turbulence present \citep{Marscher2014, Roy2023}. The degree to which the color index changes on rapid timescales can also determine whether a rapidly-varying jet angle due to precession or a wobbling jet (and therefore a changing bulk Doppler factor) is contributing to the observed variability \citep{Marchesini2016,Agarwal2016}. These discoveries, however, have taken place in detailed, observationally expensive monitoring campaigns of single objects, requiring global networks of ground-based telescopes to avoid diurnal and weather-dependent gaps. Space-based timing, especially from TESS, has enabled discoveries of rapid quasi-periodicities in the optical jet emission has led to speculation about the nature of fluid instabilities in the jet boundaries, such as the kink instability \citep{Jorstad2022,Tripathi2024a,Tripathi2024b}. However, as the TESS spacecraft has only a single wide bandpass, no color index information is present, preventing the use of these data for studying rapid inter-band lags in large samples.

\subsection{Detailed Physics of Accretion Flows}
\label{sec:accretion}

Accretion disks around supermassive black holes radiate primarily in the UV and optical, and are conceived as a series of annuli radiating like local blackbodies, decreasing in temperature with distance from the hole. This has led to enormous efforts to map accretion disks and track variations moving through accretion disks with simultaneous, multi-color modeling, to great effect. \citet{Edelson2015} and the STORM team conducted a monumental effort to monitor the Seyfert galaxy NGC~5548 with Swift and daily with HST in the X-ray, UV, and a range of optical wavebands. The results are spectacular, showing a clear signal of UV-optical lags agreeing with this general thermal annuli accretion disk theory, but with surprising results for the size of the accretion disk and with possible implications for the geometry of the broad-line region.  Further studies of individual objects with multiband, daily monitoring by STORM have led to even further insights, including detailed maps of accretion disk temperature fluctuations with both temporal and spatial resolution \citep{DeRosa2015, Cackett2023, Neustadt2024}. 

The unique insights offered by these studies cannot be overstated: this is the only direct observable into the behavior of matter in the extreme environments of AGN disks, except for direct imaging of accretion flows by, for example, the Event Horizon Telescope \citep{EHT2019,EHT2023}, or deep \emph{Chandra} imaging of gas near the Bondi radius \citep{Baganoff2003,Wong2011,Russell2015,Bambic2023}, which is only possible for a handful of nearby objects.

With cadence on minutes timescales like those afforded by TESS, but in the multiple colors explored with STORM, such insights would extend to the smallest regions of the optical disk, reaching into the most extreme space-time environments of the accreting matter. Ray tracing simulations indicate that at these scales, different wavelengths may trace the space-time environment itself \citep[e.g., ][]{Bromley1997}. 

When combined with simultaneous X-ray monitoring, multi-band timing experiments provide unique information about the geometry and location of the X-ray corona, a major contested question in AGN physics. A wealth of longer-baseline, lower-cadence X-ray / UV / optical campaigns have provided conflicting results in the question of the origin of variations: do flares initiate in the corona, which is located above the disk like a ``lamp-post", and get reprocessed successively by the accretion disk at different radii, or do flares initaite in the disk through magnetic reconnection or bulk accretion flow variations, and propagate inward to the corona?  \citep[For a recent review, see ][]{Cackett2021}.  The advent of very rapid X-ray timing with experiments like NICER \citep{Gendreau2012}, which has joint programs with TESS already, offers the possibility of simultaneous, rapid multi-band optical and X-ray experiments that probe this relationship at the smallest relevant physical scales. 

\subsection{Other Applications}
\label{sec:other}

In addition to the science cases discussed in detail here, there are numerous other applications to AGN and accretion physics. Variability has been identified as an important means of identifying potential dwarf AGN; that is, actively accreting intermediate mass black holes (IMBHs) \citep{Baldassare2020}, and indeed many low-mass AGN have been noted in TESS light curves \citep{Burke2020,Treiber2023}. Rapid, highly photometrically accurate space-based timing is well-equipped to build a census of these objects orthogonal to those found in expensive radio and X-ray imaging searches. This is important, because the occupation fraction of AGN in dwarf galaxies has major implications for the nature of black hole seeds in the early universe \citep[see ][ for a review]{Greene2020}. 

Transients also offer a tantalizing peek into accretion physics, especially those associated with tidal disruption events (TDEs) in which a massive black hole consumes a stellar object, producing a temporary accretion flow and resulting in a large, multi-wavlength flare \citep{Evans1989,Gezari2021}. TESS has seen a number of such events \citep[e.g., ][]{Holoien2019}, and has detected other interesting fast nuclear transients (including some repeating anomalous signals) in galaxies \citep[e.g., ][]{Payne2023}. Color information provides much more physical interpretability than monochrome transient detection, including characterization of the transient's evolution and information that helps constrain the nature of the disrupted star.

\section{Conclusion: The Impact of a Rapid, Multicolor Space-Based High-Energy Satellite}
\label{sec:conclusion}

Despite their original design goal of finding and determining the orbital parameters of exoplanets, the \emph{Kepler} and TESS missions' rapid optical timing with high photometric precision and uninterrupted space-based monitoring were powerful tools for exploring the high energy astrophysics of accretion and jets. Observers in these areas overcame major data reduction challenges in order to use these instruments, driven mainly by the missions' original design goals \citep[see ][ for a detailed description of challenges]{Smith2019}. Among the biggest obstacles to achieving complete pictures of the phenomena under investigation are the monolothic one-color bandpasses of the experiments and the daunting uncalibrated systematics that mimic stochastic astrophysical signals; others include the large pixel size (contributing to crowding, a major issue for faint sources) and a bright limiting magnitude that severely hampers extragalactic sample sizes. 

It is my perspective that a survey mission with many of the properties of \emph{Kepler} and TESS, but with high-energy applications in mind,  would provide major leaps forward in all of the science applications listed here. A mission with $\sim10-30$ minute cadence, multiple optical colors, and a deeper limiting magnitude, even at the expense of sky coverage (for example, limited to small, well-studied extragalactic fields) and with careful calibration geared towards recovering stochastic signal is a natural next step towards space-based high energy astrophysics in the optical regime, following an existing budget-friendly template, that would offer outstanding science returns. All extragalactic applications would benefit substantially from deeper limiting magnitudes than those probed by the exoplanet missions, due to enormous increases in sample size at magnitudes $>20$ even with reduced sky coverage. Rapid cadence offers an unprecedented window into jet and accretion disk phenomenology only accessible before with intensive global ground-based campaigns. Color information opens up a wide range of phenomenologies through the study of interband lags and leads, while also breaking degeneracies in searches for close binary AGN pairs, the source population for low-frequency gravitational waves. I urge the rest of the time domain community to consider what such a mission could do for their respective fields!

\section*{Author Contributions}

KLS is the sole author, and is responsible for all the opinions in this article, and has solely compiled the examples and references therein.

\bibliographystyle{Frontiers-Harvard} 
\bibliography{Frontiers_TDAMM_2024}


\end{document}